\documentstyle[12pt]{article}
\def\hybrid{\topmargin 0pt      \oddsidemargin 0pt
	\headheight 0pt \headsep 0pt
	\textheight 9in         
	\textwidth 6.25in       
	\marginparwidth .875in
	\parskip 5pt plus 1pt   \jot = 1.5ex}

\catcode`\@=11
\def\marginnote#1{}
\newcount\hour
\newcount\minute
\newtoks\amorpm
\hour=\time\divide\hour by60
\minute=\time{\multiply\hour by60 \global\advance\minute by-\hour}
\edef\standardtime{{\ifnum\hour<12 \global\amorpm={am}%
	\else\global\amorpm={pm}\advance\hour by-12 \fi
	\ifnum\hour=0 \hour=12 \fi
	\number\hour:\ifnum\minute<10 0\fi\number\minute\the\amorpm}}
\edef\militarytime{\number\hour:\ifnum\minute<10 0\fi\number\minute}

\def\draftlabel#1{{\@bsphack\if@filesw {\let\thepage\relax
   \xdef\@gtempa{\write\@auxout{\string
      \newlabel{#1}{{\@currentlabel}{\thepage}}}}}\@gtempa
   \if@nobreak \ifvmode\nobreak\fi\fi\fi\@esphack}
	\gdef\@eqnlabel{#1}}
\def\@eqnlabel{}
\def\@vacuum{}
\def\draftmarginnote#1{\marginpar{\raggedright\scriptsize\tt#1}}

\def\draft{\oddsidemargin -.5truein
	\def\@oddfoot{\sl preliminary draft \hfil
	\rm\thepage\hfil\sl\today\quad\militarytime}
	\let\@evenfoot\@oddfoot \overfullrule 3pt
	\let\label=\draftlabel
	\let\marginnote=\draftmarginnote
   \def\@eqnnum{(\theequation)\rlap{\kern\marginparsep\tt\@eqnlabel}%
\global\let\@eqnlabel\@vacuum}  }

\def\numberbysection{\@addtoreset{equation}{section}
	\def\theequation{\thesection.\arabic{equation}}}

\def\underline#1{\relax\ifmmode\@@underline#1\else
	$\@@underline{\hbox{#1}}$\relax\fi}

\def\titlepage{\@restonecolfalse\if@twocolumn\@restonecoltrue\onecolumn
     \else \newpage \fi \thispagestyle{empty}\c@page\z@
	\def\thefootnote{\fnsymbol{footnote}} }

\def\endtitlepage{\if@restonecol\twocolumn \else  \fi
	\def\thefootnote{\arabic{footnote}}
	\setcounter{footnote}{0}}  
\catcode`@=12
\relax

\def\dd{d+s}
\def\ll{\lambda^\t}
\def\d{\delta}
\def\trois{{3\over 2}}

\def\ba{{\bar\alpha}}

\def\ee{\eea}
\def\be{\bea}

\def\eq{\eeq}
\def\bq{\beq}
\def\ie{\hbox{\it i.e.}}
\def\etc{\hbox{\it etc.}}
\def\eg{\hbox{\ite.g.}}
\def\cf{\hbox{\it cf.}}

\def\etal{\hbox{\it et al.}}
\def\tr{\mathop{\rm tr}}
\def\Tr{\mathop{\rm Tr}}
\def\beq{\begin{equation}}
\def\eeq{\end{equation}}
\def\bea{\begin{eqnarray}}
 \def\eea{\end{eqnarray}}

\def\bar{\overline}
 \def\z{{\bar {z}}}
 \def\nn{\nonumber}
\def\pa{\partial}
\def\d{{\cal D}}

\def\t{{\theta}}
\def\s{\sigma}

 \def\pt{\partial_\t}
\def\ic{\int_0^T}

\def\vq{\vec q}

\def\P{\Psi}
\def\bP{\bar \Psi}

\def\vP{\vec \P}
\def\vbP{\vec {\bP}}

\def\l{\lambda}
\def\vl{\vec \l}

\def\eij{\epsilon^{ij}}

\def\V{{\dot q}_i+{{\delta V}\over{\delta q_i}}}
\def\VP{{\dot \P}_i+{{\delta ^2 V}\over{\delta q_i \delta q_j}}\P_j}

\def\v{{\dot q}_i+f{{\eij q_j}\over{\vec q^2}}}

\def\p{\vec p}

\def\demi{{1\over 2}}

\relax
\numberbysection
\hybrid
\begin{document}

 \begin{titlepage}
 \hfill Paris LPTHE 94--05
\vskip 38 truemm
\begin{center}

{\large\bf
FROM TOPOLOGICAL QUANTUM MECHANICS
TO THE PHYSICS OF SPIN-ONE PARTICLES \\}
{Romain Attal and Laurent Baulieu \footnote{e-mail address:
attal@lpthe.jussieu.fr \quad baulieu@lpthe.jussieu.fr}}

 {\it LPTHE Universit\'es Pierre et Marie Curie et Denis Diderot
\\Laboratoire associ\'e au CNRS, URA 280\\
4 place Jussieu, Tour 16, 1er \'etage

 F-75252 Paris Cedex 05, France.}\\

\end{center}

\vskip 1cm

\begin{quotation}
{\bf Abstract: }
 We give  an example of topological theory whose Hilbert
space contains physical objects: the $N=2$ supersymmetric
Lagrangian for spin-one particles moving in $D$-dimensional
space-time is a topological model in a space with two more
dimensions. The equivalence is still valid in a curved space-time.
As an application, we calculate the deviation of spin-one particles
in a Schwarzschild background.
\end{quotation}
\end{titlepage}

\newpage

\def\dd{d+s}
 \def\ll{\lambda^\t}
\def\d{\delta}
\def\trois{{3\over 2}}

\def\ba{{\bar\alpha}}

\def\ee{\eea}
\def\be{\bea}

\def\eq{\eeq}
\def\bq{\beq}
\def\ie{\hbox{\it i.e.}}
\def\etc{\hbox{\it etc.}}
 \def\eg{\hbox{\ite.g.}}
  \def\cf{\hbox{\it cf.}}

 \def\etal{\hbox{\it et al.}}
\def\tr{\mathop{\rm tr}}
 \def\Tr{\mathop{\rm Tr}}
\def\beq{\begin{equation}}
\def\eeq{\end{equation}}
\def\bea{\begin{eqnarray}}
 \def\eea{\end{eqnarray}}

\def\bar{\overline}
 \def\z{{\bar {z}}}
 \def\nn{\nonumber}
\def\pa{\partial}
\def\d{{\cal D}}
\def\L{{\cal L}}
\def\t{{\tau}}
\def\s{\sigma}
  \def\pt{\partial_\t}
\def\ic{\int_0^T}
\def\a{{\alpha}}
\def\vq{\vec q}

\def\P{\Psi}
\def\bP{\bar \Psi}

\def\vP{\vec \P}
\def\vbP{\vec {\bP}}

\def\l{\tau}
\def\vl{\vec \l}
\def\p{  \varphi}

\def\eij{\epsilon^{ij}}
\def\v{{\dot q}_i+{{\eij q_j}\over{\vec q^2}}}
\def\V{{\dot q}_i+{{\delta V}\over{\delta q_i}}}
\def\VP{{\dot \P}_i+{{\delta ^2 V}\over{\delta q_iq_j}}\P_j}

\def\nn{\nonumber}
\relax
\hyphenation{con-ti-nu-um}
\def\re{reparametrization}

\def\PP{{\P^{D+1}}}
\def\dPP{{\dot\P^{D+1}}}
\def\hPP{{\hat\P^{D+1}}}
\def\bPP{{\bar\P^{D+1}}}
\def\m{\mu}
\def\n{\nu}
\def\gmn{g_{\m\n}}
\def\G{\Gamma}
\def\r{\rho}
\def\e{\alpha}

 \section{Introduction.}
 Supersymmetric quantum mechanics  has   specific
applications: it can be used to describe the dynamics of spinning point
particles \cite{berezin}
as well as to compute a certain number of topological invariants of the
target space \cite{Witten}. Moreover it
gives an  insight into the structure of superstring theory,
and has opened the way to its modern
formulation \cite{brink} \cite{polyakov}.

In this paper  we point out  an example showing that
topological quantum
theories may exhibit  a  phase with a  Hilbert space made of
particle degrees of freedom. We implement the idea that the fundamental
symmetries of nature could be originally of the topological type
and that the  observed gauge symmetries would be obtained  by gauge-fixing
the huge topological
symmetry in a   BRST invariant way, leaving therefore
an  $N=2$ supersymmetric theory.

Our example is the description of spin-one particles by a
$N=2$ supersymmetric action. The use of anticommuting variables to
describe spinning particles was introduced in \cite{berezin}.
Then, it was found  that
 local supersymmetry  of rank $2S$  on the
worldline is necessary to describe consistently a particle of spin $S$.
The resulting      constrained system \cite {dirac} \cite{siegel}
  requires a careful
gauge-fixing of the einbein  and the gravitini. One  obtains eventually a
tractable  Lagrangian   formulation \cite{polbook}, \cite{vh}.
(There are many references on the subject, of which we quote very few).

In this work we interpret   local   supersymmetry   on the worldline
as a residue of a more
fundamental topological symmetry, defined in a target-space with two extra
dimensions. The word ``topological'' has to be understood in the
sense that one does the BRST invariant gauge-fixing of a classical action,
such   as a characteristic number of the base manifold,
which is invariant under arbitrary infinitesimal deformations
of the fields \cite{review}.
One of the coordinates is eventually identified as the einbein on the
worldline. Other fields  must be introduced to
enforce the topological BRST invariance. They can be eliminated by their
equations of motion and decouple from the physical sector.
To obtain in a natural way  a nowhere vanishing  einbein, we
use a disconnected   higher dimensional target-space where the
hyperplane $\{e=0 \}$ is a priori extracted. Thus, one  introduces
some topology before any gauge-fixing. Two disconnected topological
sectors exist, $\{e>0 \}$ and $\{e<0 \}$, which correspond to the
prescription $\pm i\epsilon $ for the  propagators. It is fundamental that
the gauge functions be compatible with the topology of space: they must
induce a potential which rejects the trajectories from the   hyperplane
 $\{e=0 \}$.

The paper is organized as follows. We first review the
supersymmetric description
of a relativistic spinning particle in a Riemannian  space-time. Then
we consider the    case  of $N=2$ supersymmetry and show our main
result: the  link between the supersymmetric description of scalar
or spin-one particles and topological
quantum mechanics in a higher dimensional target-space. In the last section, we
verify that the    constraints of the theory    identify its   physical
content and illustrate the result by computing the deviation of the
trajectories from geodesics due to the interactions between  geometry  and
spin.

\section {  Worldline supergravity  for  a spinning particle and its gauge
fixing.}

 Consider a spin-$S$ particle in a $D$-dimensional   space-time.
Classically, it follows a worldline whose coordinates    $ X^\m(\l)$ are
parametrized by a real number  $\l$. If the particle is massive, a natural
choice of this parameter is the proper-time. The idea originating from
\cite{berezin}
 is to  describe  the spin of the particle by assigning to each
value of $\l$ a vector with anticommuting coordinates ${\P}_i^\m(\l)$ where the
vector index $\mu$ runs between $1$ and $D$ and $i$ between  $1$ and $2S$.
Indeed, in the case of a flat space-time  and spin one-half,
the Lagrangian   density introduced in
\cite{berezin}    is
\be
\L=\demi(\dot X^2(\l)-\P^\m(\l)\dot\P_\m(\l))
\ee
where the dot $\dot { }$ means $\pt$, $\t$ being a parametrization of
the worldline.
Upon canonical quantization $\P^\m(\l)$ is replaced
by a $\l$-independent operator  $\hat \P^\m $ which satisfies anticommutation
relations
\be \label{com}
\{\hat \P^\m , \hat \P_\n\}_+=2\delta^\m_\n
\ee
The   Hamiltonian is
\be
H=\demi p^2=\demi Q^2
\ee
with $Q=p_\m\hat\P^\m$.
Due to (\ref{com}) the $\hat \P$'s can be represented by Dirac matrices
and $Q$ is the free Dirac operator. $Q$ commutes with $H$ and it makes
sense to consider the restriction of the Hilbert space to the set of states
$|\p>$ satisfying  \be\label{dirac}
Q|\p>=0
\ee
By definition of $Q$, this equation means that  the $|\p>$ are the states of a
massless spin one-half particle. The extension to the case of a massive
particle implies the
introduction  of an additional Grassmannian variable $\PP$
and the generalization of  $\L$ to
\be\label{lagrangiengl}
\L=\demi(\dot X^2(\t)-\P^\m(\t)\dot\P_\m(\t)
-\PP(\t)\dPP(\t)+m^2)
\ee
(Formally, $\dot X ^{D+1} \to m $), so that
\be
H=\demi (p^2-m^2)=\demi Q^2
\ee
with
\be
Q=p_\m\hat\P^\m+m{\hat\Psi}^{D+1}
\ee
and one has in addition to (\ref{com})
\be \label{comm}
\{\hPP , \hPP\}_+=-2 \quad \quad
\{\hat \P^\m , \hPP\}_+=0
\ee
The condition (\ref{dirac}) is now the free Dirac equation for a spin one-half
particle of mass $m$, multiplied by $ \hPP$.
The generalization to the case of an arbitrary spin is obtained by
duplicating  $2S$ times the components of
$\P$, $\P^\m \to \P^{\m}_i, 1\leq i\leq 2S$, as can be
seen by constructing the  representations of $SO(D)$ by
suitable tensor products of spin one-half representations
\cite{townsend}   \cite{goteborg}.

To understand the constraint (\ref{dirac}), it is in fact necessary to promote
the global supersymmetry of the action, corresponding to the commutation of
$H$ and $Q$, into a local supersymmetry. Indeed, when   time
flows, the state of the particle must  evolve from a  solution of
the Dirac equation to another solution of this equation, without any
possibility to collapse in an unphysical state (out of $Ker(Q)$). A natural way
to reach such a unitarity requirement is to impose
the supersymmetry   independently for all values of $\t$, that is, to gauge the
supersymmetry on the worldline. In this way,   the condition (\ref{dirac})
appears as
the   definition of physical states in a gauge
theory with generator $Q$ which ensures unitarity,  like the transversality
condition of gauge bosons in ordinary Yang-Mills theory.
 For
consistency,    the  diffeomorphism invariance on the worldline must be also
imposed
 since the commutator of two supersymmetry  transformations contains
a diffeomorphism. One thus introduces gauge fields for these symmetries,
the einbein $e(\t)$ and the (anticommuting) gravitino  $\a(\t)$.
By minimal coupling on the worldline, (\ref{lagrangiengl}) is thus
generalized to  the following Lagrangian which is locally supersymmetric
and reparametrization invariant, up to a pure derivative
with respect to  $\t$ :
\be \label{lagrangienl}
\L=\demi \left( e^{-1} {\dot X}^2
-\P(\dot \P + \a e^{-1} \dot X )
-\PP(\dPP + m\a )
+em^2 \right)
\ee
(In this section we omit the vector and spin indices).
Formally, $\dot X ^{D+1} \to me$.
The transformation laws of $e$ and $\a$
are those of one-dimensional supergravity of rank $2S$.

The gauge-fixing $e(\t)=1$ and $\a(\t)=0$
identifies (2.5) and (2.9), up to Faddeev-Popov ghost terms.
These ghost terms have a supersymmetric  form
$b\dot c+\beta \dot\gamma$. They decouple effectively, since their effect is to
multiply all the amplitudes by a ratio of determinants, independent
of the metric in space-time.
This gauge-fixing  is however inconsistent because it is too strong, since
the Lagrangian   is gauge invariant only up to boundary
terms. Therefore, given a general gauge transformation, one must  put
restrictions on its parameters to get the invariance of the action, and there
are  not enough degrees of freedom in the symmetry to enforce the gauge
 $e(\t)=1$ and $\a(\t)=0$. One can at most set  $e(\t)=e_0$ and $\a(\t)=\a_0$,
letting the constants $e_0 >0$ and $\a_0$ free, that is,
doing an ordinary integration over $e_0$ and   $\a_0$ in the path integral
after the gauge-fixing \cite{polbook}.
This yields the following partition function for the theory
\be\label{pi}
Z=\int_0 ^{\infty} de_0 \int d\a_0
\int[d X(\t)][d\P(\t)]\exp -\int_0 ^1 d\t \L_0
\ee
with
\be
\label{lagrangienl0}
 \L_0=\demi \left( {e_0}^{-1}
{\dot   X}^2 +e_0 m^2
-\P (\dot\P + \a_0 {e_0}^{-1} \dot X)
-\PP(\dPP + m \a_0)\right)
\ee

Using the Lagrangian  (\ref{lagrangiengl}) instead of
(\ref{lagrangienl0}) implies that one misses crucial spin-orbit interactions
described by the Grassmannian integration over the constant $\a_0$ which
induces the fermionic constraint
 $ \int d\t(\P \dot X+me_0 \PP) =0$. The use of (\ref{lagrangiengl})
leads indeed to a spin-zero particle
propagator while (2.11)     leads to the expected  spin one-half
propagator.  One gets the $\pm i\epsilon$ propagators
depending on the choice of    the integration domain $\{e_0 >0 \}$ or
$\{e_0 <0 \}$. Notice that the $e$-dependence of the Lagrangian
(\ref{lagrangiengl}) gives a negligible weight in the path integral
(\ref{pi}) to  the trajectories    with points near the hyperplane
    $\{e_0 =0\}$.
The integration over $e_0$ and $\a_0$ has
a simple interpretation  in Hamiltonian formalism. The Hamiltonian associated
to (\ref{lagrangienl0}) is
\be
H&=&{e_0 \over 2}(p^2-m^2)+{\a_0 \over 2}(p_\m \hat\P^\m+m\hPP)\nn\\
&=&{e_0 \over 2}(p^2-m^2)+{\a_0 \over 2} Q
\ee
The constants $e_0$ and $\a_0$ are thus
Lagrange multipliers which force the particle to satisfy   the Klein-Gordon
equation  and the  Dirac equation  (or its higher spin generalizations
$Q_i|\p>=0 $). Observe that in Lagrangian   formalism, the Klein-Gordon
equation  is not a consequence of the Dirac equation, due to the
anticommutativity of Grassmann   variables, and the two constraints
$Q|\p>=0$ and $H|\p>=0$ must be used separately. Therefore, we have  a theory
where the Hamiltonian is a sum of   constraints, which leads to
known technical difficulties \cite{dirac}\cite{siegel}.
In Lagrangian   formalism, supergravity on the worldline and its
correct gauge-fixing take care of all details \cite{polbook}.

 The above description is valid for a flat space-time. It
can be  generalized to the case where the
particle moves in a curved space-time  and/or couples to an
external electromagnetic field, by minimal coupling in the target-space.
The compatibility between the worldline diffeomorphism
invariance and local supersymmetry with reparametrization invariance
in the target-space for a general metric $\gmn$ is however possible only for
 $N\leq 2$ \cite{townsend}. This phenomenon is possibly   related to the
limited number of consistent supergravities \cite{petervan}.

We now consider the case $N=2$ and establish our main result in this paper: the
link of the theory with a topological model.

\section{The N=2 supersymmetric action as a topological action.}

The $N=2$ supersymmetric Lagrangian  with a general
background metric
$\gmn$ is
\be\label{lone}
\L _{SUSY}&=& {1 \over 2e } \gmn  \dot X^\m
\dot X^ \n
-\bP^\m ( \gmn \dot\P^\n +e\G_{\m\n\r}\dot X ^\n \P ^\r)
 +e^{-1} \gmn \dot X^\m (\bP^\n \a+\ba \P^\n )
\nn\\
&+&{e m^2 \over 2}-\bPP  \dPP + m (\bPP \a + \ba \PP) \nn \\
&-&e^{-1} \ba \a \bP \P
+{e \over 2 } R_{\m\n\r\s}\bP^\m\P^\n\bP^\r\P^\s
 \ee
where $\P $ and $\bP$ are independent Grassmannian coordinates.
(Compare with \cite{townsend}).
The Lagrangian (\ref{lone}) has two local supersymmetries, with generators $Q$
and $\bar Q$.  An $O(2)$ symmetry between
$\P $ and $\bP$ can be enforced by introducing a single gauge field $f(\t)$
and adding a term $f \bP \P  $.
However, no new information is provided, since   one increases the
symmetry by one generator, which is  compensated by the introduction
of the additional  degree of freedom carried by $f$.
The latter can indeed  be gauge-fixed to zero and one recovers
(\ref{lone}). Moreover, in view of identifying
$\P$ and $\bP $ as ghosts and antighosts, one wishes    to freeze the
symmetry between these two fields.  We thus ignore the possibility of gauging
the  $O(2)$ symmetry. We will check in the next
section that the  Hilbert space associated to the Lagrangian
(3.1) contains spin-one particles.

The Lagrangian  (\ref{lone}) can be conveniently rewritten
in first order formalism by introducing a Lagrange
multiplier $b^\m(\t)$. One gets the equivalent form
\be
\label{ltwo}
\L_{SUSY} \sim -{e\over 2  }(\gmn b^\m b^\n-m^2)
 +\gmn b^\m (\dot X^\n  +e\G^\n_{\r\s}\bP^\r\P^\s+\bP^\n \a+\ba \P^\n)\nn\\
-\bP^\m (\gmn \dot\P^\n +e\pa_\r g_{\m \n} \dot X^\n \P^\r)
 -\G_{\n \r \s} \bP^\r \P^\s(\bP^\n \a+\ba \P^\n )
\nn\\
-\bPP \dPP +m(\bPP \a +\ba\PP)-{e \over 2}
\pa_\n\G_{\m\r\s}\bP^\m\P^\n\bP^\r\P^\s
\ee
(The symbol $\sim$ means that the  two Lagrangians differ by a term
which can be eliminated using an algebraic equation of motion,
and, consequently,
define the same quantum theory).
For $e=1$, $\a=\ba=0$ and $\PP=\bPP=0$,  the Lagrangian  (3.2)
can be interpreted as the gauge-fixing of zero or of a term invariant under
isotopies of the curve $X$ \cite{baulieusingermq}. In this interpretation
the $\P$ are topological ghosts and the $\bP$ are antighosts. The BRST
graded differential operator $s$ of the
topological symmetry is defined by
\be
sX^\m =\P^\m\nn\\
s\P^\m =0\nn\\
s\bP^\m =b^ \m \nn\\
sb^\m =0
\ee
and the gauge-fixing Lagrangian is $s$-exact
 modulo a pure derivative
\be
\L_{GF}=s(\bP_\m(-\demi b^\m+\dot X^\m  +{1 \over 2}\G^\m_{\r\s}\bP^\r \P^\s))
\ee
(Since $s^2=0$, $\L_{GF}$ is s-invariant.)
To identify (3.1) as a topological Lagrangian, we must introduce new
ingredients. We will enlarge the
target-space with two additional components, and add a ghost of ghost.
We will eventually identify one  of the extra coordinates  with the einbein $e$
and the other one
will be forced to vary in a Gaussian way around an arbitrary scale,
with an arbitrary width.
The gravitini $\a$ and $\ba$ of the effective worldline supergravity
will be interpreted as ghosts of the topological symmetry. The  $O(2)$
invariance  corresponds to the ghost number conservation.

  We consider a $(D+2)$-dimensional space-time with coordinates
$X^A=(X^\m,X^{D+1}=e, X^{D+2})$. We exclude from the space the hyperplane
$\{X^{D+1}=0 \}$ which yields two separated half-spaces, characterized by   the
value of $sign (e)$.
  We wish to define a partition function through a
path integration over the curves $X^A(\t)$, with a topological action
which is invariant under the BRST symmetry
associated to isotopies of this curve in each half-space.
In other words we wish to construct an action by consistently
gauge-fixing the topological Lagrangian  $sign (e)$.
In a way which is analogous to the case of
topological Yang-Mills symmetry, where one gauge-fixes the second
Chern class $\int Tr \ F^2$ \cite {bs},  we  combine the pure topological
symmetry, with topological ghosts $\P^A_{top}(\t)$, to the diffeomorphism
symmetry on the curve, with Faddeev-Popov ghost $c(\t)$.
The apparent redundancy in the number of ghost variables
$\P^A_{top}(\t)$ and $c(\t)$, which exceeds the number of bosonic
classical variables, is  counterbalanced by the
introduction of a ghost of ghosts $\Phi(\t)$ with ghost number two.
The action of the BRST differential $s$ is defined by
\be
sX^\m&=&\P^\m_{top}+c\dot X^\m  = \P^\m
  \nn\\
se &=&\P^e_{top}+c\dot e  = 2 \eta =\a+\dot\P^{D+1}
 \nn\\
sX^{D+2}&=&\P^{D+2}_{top}+c\dot X^{D+2} = \P^{D+2}
\nn\\
s\P^\m &=&0
\nn\\
s\P^{D+2}&=&0
\nn\\
s\P^{D+1}&=&\Phi
\nn\\
s\a&=&-\dot\Phi
\nn\\
s\Phi&=&0
\ee
In agreement with  the art of BRST invariant
gauge-fixing, we introduce $D+2$ antighosts with ghost
number $(-1)$ and the associated Lagrange multipliers for the
gauge conditions on the $X^A$'s. We also introduce an antighost
$\bar\Phi$ with ghost number $(-2)$ and its fermionic partner $\bar\eta$
with ghost number $(-1)$ which we will use as a fermionic
Lagrange multiplier for the gauge condition in the ghost sector.
In this sector the action of $s$ is
 \be
s\bP^A &=&b^A
\nn\\
 sb^A &=&0
\nn\\
s\bar\Phi &=& \bar\eta\nn\\
s\bar\eta &=&0
\ee
The gauge-fixing Lagrangian    must be written as an s-exact term
\be
\L^X+\L^{D+1}+\L^{D+2}+\L^\Phi=s\left(\bP^A(\ldots)_A+\bar\Phi(\ldots)\right)
\ee
For the gauge-fixing in  the $X$-sector, we choose
\be\label{lx}
\L^X=
&s&\left( -{e\over 2}\gmn \bP^\n b^\m+
\gmn\bP^\m(\dot X^\n
+\bar\eta\P^\n
+{e \over 2}\G^\n_{\r\s}\bP^\r\P^\s)\right)\nn\\
=&-&{e \over 2  } \gmn b^\m b^\n
 +\gmn b^\m
(\dot X^\n  +e\G^\n_{\r\s}\bP^\r\P^\s
+\bP^\n \eta
+\bar\eta\P^\n )\nn\\
&-&\bP^\n (\gmn\dot\P^\m +e\pa_\r\gmn\dot X^\n \P^\r)
-{e \over 2 } \pa_\n\G_{\m\r\s}\bP^\m\P^\n\bP^\r\P^\s
- \G_{\m\r\s}\bP^\m \eta \bP^\r \P^\s\nn\\
 \ee
For the gauge-fixing in the $e$-sector, we choose
\be
\L^{D+1}=
-s\left(\bPP e (m+{b^{D+1} \over 2})\right)
=-e{(b^{D+1})^2 \over 2}+
b^{D+1}(-me+\bPP\eta)+2m\bPP \eta \nn \\
\ee
After   elimination of the field
$b^{D+1}$,we obtain
\be\label{ld1}
\L^{D+1} \sim
{e m^2
\over 2}+m \bP^{D+1} \eta
\ee
For the gauge-fixing in the $X^{D+2}$-sector, we choose
\be
\L^{D+2}&=&
s\left(\bP^{D+2}\left(- {a \over 2}  b^{D+2}+X^{D+2}-C-{1 \over a}
{{\bP^{D+1}\dot{\P}^{D+1}}\over{X^{D+2}-C}}\right) \right)
\nn\\
&=&  -{a \over 2} (b^{D+2})^2+b^{D+2}
\left(X^{D+2}-C-a{{\bP^{D+1}\dot{\P}^{D+1}}\over{X^{D+2}-C}}\right)\nn \\
& &
-\bP^{D+2}\left(\P^{D+2}-a s\left({{\bP^{D+1}\dot{\P}^{D+1}}\over
{X^{D+2}-C}}\right)\right)
\ee
$a$ and $C$ are arbitrarily chosen real numbers. After elimination of the field
$b^{D+2}$, we find
\be
\L^{D+2}
\sim
-\bP^{D+1}\dot{\P}^{D+1}
+
{1 \over 2a}(X^{D+2}-C)^2
-
\bP^{D+2}\left(\P^{D+2}-a s\left({{\bP^{D+1}\dot{\P}^{D+1}}\over
{X^{D+2}-C}}\right) \right)\nn\\
 \ee
The variable $X^{D+2}$ can be eliminated by its algebraic equation of motion as
well as
the corresponding ghosts  $\P^{D+2}$ and $\bP^{D+2}$, after some field
redefinitions. $X^{D+2}$ is
concentrated in a Gaussian way
around the arbitrary scale $C$, with an arbitrary width $a$.
We are thus left with   the propagating term for $\P^{D+1}$ and
$\bP^{D+1}$  which was
missing in $\L^X$ and $\L^{D+1}$
\be\label{ld2}
\L^{D+2}
\sim
-\bP^{D+1}\dot{\P}^{D+1}
 \ee
We finally choose the gauge-fixing  in the ghost sector.
To recover the full Lagrangian      (\ref{ltwo}) and eventually identify the
coordinate $e$ as the einbein of the projection of the particle trajectory in
the $D$-dimensional physical space-time, we need a term linear
in $\bar\eta$ as well as another term to get rid of unwanted higher
order fermionic terms. We   define
\begin{eqnarray}
\L^{\Phi}&=&s(\bar\Phi(m \P^{D+1}
- \G_{\n\r\s} \bP^\r \P^\s \P^\n )) \\ \nn
&=&\bar\eta(m\P^{D+1}
- \G_{\n\r\s} \bP^\r \P^\s \P^\n )
+\bar\Phi(m \Phi-s(\G_{\n\r\s} \bP^\r \P^\s \P^\n ))
\nn
\end{eqnarray}
The dependence on the ghosts of ghosts $\Phi$ and $\bar\Phi$ is trivial: these
fields decouple after a Gaussian integration.
One has thus
\be\label{lfi} \L^{\Phi}\sim
m \bar\eta \P^{D+1} - \G_{\n\r\s} \bP^\r \P^\s \bar\eta \P^\n
\ee
Adding all terms (\ref{lx}), (\ref{ld1}), (\ref{ld2}) and (\ref{lfi}),
we finally recognize that
$\L^X+\L^{D+1}+\L^{D+2}+\L^\Phi$ is equivalent to
the Lagrangian (3.2), modulo the elimination of auxiliary fields and the
change of notation $(\eta,\bar\eta) \to (\alpha,\bar\alpha)$.
We have therefore shown the announced result:
the $N=2$ local supersymmetry of the Lagrangian describing spin-one particles
is a residual symmetry coming from a topological model
after a suitable gauge-fixing.

\section{Derivation of Proca's equations from the (N=2)-supersymmetric action.}

To verify the physical content of the model of the last section, we consider a
flat space-time, and choose the   gauge where the einbein and gravitini
are constants over which we integrate. As explained in the first
section, the Hamiltonian is
 \be
H={e_0 \over 2}(p^2-m^2)+\ba_0 Q +\a_0\bar Q
\ee
with
\be
Q=p_\m \P^\m+m \PP
\nn\\
\bar Q = p_\m \bP^\m+m \bPP
\ee
The matrices $\P$ and $\bP$ satisfy the Clifford algebra
\be
\{\P^A,\bP^B\}_+=\eta^{AB}\quad , \quad
\{\P^A,\P^B\}_+=\{\bP^A,\bP^B\}_+=0
\ee
for $A,B=1,...,D+1$.
Since the underlying gauge symmetry has $Q$ and $\bar Q$ as
generators, the   physical states
 satisfy
\be\label{proca}
 Q|\phi>=0 \quad\quad \bar Q |\phi>=0
\ee
in addition to
\be
(p^2-m^2)|\phi>=0
\ee
The $\P$ and $\bP$ are generalizations of the Pauli matrices, and it is
convenient to use a Schwinger type construction, in order to exploit directly
their Clifford algebra structure.
One introduces a spin vacuum
$|0>$ annihilated by the $\P$'s. Then, the $\bP$'s can be identified as their
adjoints and act as creation operators. In the $X$ representation, we can
write
a general state as
\def\bp{\bar\p}
\be
|\phi>= \left(\p_0+\p_\m\bP^\m +\p_{\m_1\m_2}\bP^{\m_1} \bP^{\m_2}
+\ldots+
\p_{\m_1\ldots\m_D}\bP^{\m_1} \ldots \bP^{\m_D} \right) |0>
\nn\\
+ \bPP \left(\bp_0 +\bp_\m\bP^\m  +\bp_ {\m_1 \m_2}\bP^{\m_1} \bP^{\m_2}
+\ldots+
\bp_{\m_1\ldots\m_D} \ldots \bP^{\m_D}
\right)
  |0>
\ee
The wave functions $\p_{\m_1\ldots\m_p}(X)$ and $\bp_{\m_1\ldots\m_p}(X)$ are
antisymmetric and it is useful to consider the differential forms
\be
\p_p={1\over{p!}}dX^{\m_1}\ldots d X^{\m_p}\p_{\m_1\ldots\m_p}(X)
\nn\\
\bp_p={1\over{p!}}dX^{\m_1}\ldots d X^{\m_p}\bp_{\m_1\ldots\m_p}(X)
\ee
for $0\leq p\leq D$.
The constraints (\ref{proca}) can be conveniently written as
\be
d\p_p+im\bp_{p+1}=0
\ee
\be
d^*\bp_p+im\p_{p-1}=0
\ee
\be
d\bp_p=0
\ee
\be
d^*\p_p=0
\ee
Where $d=dx^\m\pa_\m$ and $d^*$ is its Hodge dual.
One has also
\be
(d^* d + d d^*)\p=-m^2\p\quad \quad (d^*d+d d^*)\bp=-m^2\bp
\ee
These equations determine the independent degrees of
freedom. When $m\neq 0$, they couple the two sectors of opposite
chiralities. Moreover, when $D$ is even, the first one contains
${D\over2}$  forms, namely one   scalar $(\p _0)$, one vector $(\p_1)$,
..., and one $({D\over 2}-1)$-form  ($ \p_{{D\over 2}-1} $).
The other one has a dual structure ($ \bp_{{D\over 2}+1} $,...,$\bp_D$).
For $(\p_1)$, the constraints (4.4) can be rewritten:
\be
\pa_\m \bp^{\m\n}+im\p^\n=0
\nn\\
\pa_\m\p^\m=0
\nn\\
\pa_{[\m}\p_{\n]}+im\bp_{\m\n}=0
\ee
Thus the vector wave function $\p_1$ satisfies Proca's equations,
and descibes a spin-one particle with mass $m$.
It follows that the field equations of $\p_1$ and $\bp_1$ can
be derived by minimizing Proca's Lagrangian
\be
\L_{Proca}
={m\over 2}\bar\p_{\m\n}\p^{\m\n}-{i \over 2}\p^{\m\n}(\pa_\m\p_\n-\pa_\n\p_\m)
\nn\\
-{i \over 2} \bp^{\m\n}(\pa_\m\bp_\n-\pa_\n\bp_\m)
+m\bp_\m\p^\m
\ee
When $m=0$, the two sectors of opposite chiralities decouple.
In each sector, the independent degrees of freedom are now one 0-form
$A_0$ (with $\p_1=dA_0$), one   1-form $A_1$ (with $\p_2=dA_1$),...,
one (D-2)-form $A_{D-2}$ (with $\p_{D-1}=dA_{D-2}$).
The $\p_p$'s  are closed and co-closed, i.e. the $A_{p}$'s
satisfy Maxwell's equations and are defined up to gauge
transformations. Consequently, $\p_2$ can be identified with the
field strength of a photon.
If we consider the case $D=4$ and $m\neq 0$, the spectrum reduces
to two scalars and two massive spin-one particles, and contains
8=2(1+3) degrees of freedom.
For  $m=0$, we have two massless scalars and two massless vectors, so
that we still have 8=2(1+1+2) independent degrees of freedom.

As an application of this formalism, we   study the classical behavior of
spinning particles in a curved space-time. We are interested in the
approximation  where the trajectory of the particle is classical, while  the
spin effects are visible as it would be  the case in  a Stern-Gerlach
experiment.
This situation occurs if  the order of magnitude of the
interaction energy between the spin and the curvature, which is essentially
proportional to the space-time curvature times $\hbar$ (analogously to
the interaction between
 the the magnetic field and a magnetic moment due to the spin), is comparable
to the kinematical energy of the particle. One must also    measure
the position of the particle on a domain much larger than its Compton
wavelength.  In this limit
 the
position $X^\m$   and momentum
 $P_\m$ are   ordinary numbers and the quantum Hamiltonian becomes simply a
matrix built from the $\P$'s and $\bP$'s acting in the
spin-space with coefficients depending on the classical position $X$ and
 momentum $P$.
 The $\t$-dependence of the classical dynamics of the particle can be expressed
by applying Hamilton-Jacobi's method with this matricial
Hamiltonian.  The only  quantum effects are due to the spin
interaction with the space-time curvature.
(In a fully classical approximation,
$\hbar=0$, and the spin effects disappear, since all the fermionic
operators are proportional to $\sqrt \hbar$.)
One can always find a basis for the
spin states, which depends on the space-time position and such that the
Hamiltonian is diagonal. In this basis the spin value is conserved
through   evolution, i.e.
the spin observables are   paralelly transported  along the trajectory.
By diagonalization in spin space,  $H$ determines
independent
Hamilton-Jacobi's equations for each spin degree of freedom of the particle.
For the spin-one case, we expect three different trajectories corresponding to
the values 1, 0 and -1 for the projection of the spin on a spatial axis
in the rest frame of the particle.

We  consider  the case of a Schwarzschild gravitational field in four
dimensional space-time
($ds^2=(1-{r_0 \over r})dt^2-(1-{r_0 \over r})^{-1}
dr^2-r^2 (d\theta^2+sin^2 \theta \ d\phi^2)$
with
$r_0=2GM/c^2$.)
We will compute the correction, due to the spin, to
Einstein's formula predicting the shift of the perihelion of a
spinless point particle. For the other classical test of general
relativity, i.e. the bending of light rays in a gravitational field, we will
find that the wave vector of a polarized photon deviates from
geodesic motions by a relative shift proportional to $ \hbar $. These results
are in agreement
with the fact that  a particle with an angular momentum interacts
with the space-time curvature, as first pointed out by Papapetrou for a
rotating body \cite{Papapetrou}.
The advantage of a supersymmetric Hamiltonian is that it
defines unambiguously the spin effects.
Since we work to first non-trivial order in $ \hbar $,  we restore
from now on  the
$ \hbar $ dependence in the formulae.
The matricial Hamilton-Jacobi's equation is  obtained by replacing
in the supersymmetric Hamiltonian  the classical momentum $p_\m$
by ${\delta S \over \delta X^\m}$ where
$S[X^\m,\t]$ is the   action of the classical trajectory of the particle in
a given spin state, with arbitrarily chosen initial and final boundary
conditions. Notice that keeping the lowest order in $\hbar$ means
that we only retain the covariant derivative of
the fermionic variables and not the curvature term. This yields
\be\label{hj}
g^{\m\n} {\delta S \over \delta X^{\m}}
{\delta S \over \delta X^{\n}}
-m^2+2\hbar {\delta S \over \delta X^{\m}} \omega^{\m}_{ab}
\Sigma^{ab} +O(\hbar^2) =0\ee

The space-time spin-connection $\omega$ is related to the space-time vierbein
$E$ and to Christoffel's symbol $\Gamma$
\be
\omega_{\m ab}=E^{\a}_a {E^{\beta}_b} {\Gamma_{\a\m\beta}} \\
E_{\a}^a E^b_{\beta} \eta_{ab}=g_{\a\beta} \\
\Gamma_{\a\m\beta}=\demi(\partial_{\m} g_{\a\beta}+\partial_{\beta} g_{\a\m}
-\partial_{\a} g_{\beta\m})
\ee
The $\Sigma ^{ab} ={i \over 2}\left(\bP^a \P^b - \bP^b \P^a \right)$
are the generators of the
(reducible) 32-dimensional  representation of the Lorentz group  defined by the
algebra (4.3) and acting on the states solving (4.6).
If the matrix form of $\bP^5$ is
chosen diagonal, the spin operators  $\Sigma ^{ab}$ become block-diagonal
with two independent sectors of opposite chiralities, corresponding to the
eigenvalues $0$ and $1$ of $\bP^5\P^5$, so the
32-dimensional  representation splits into two independent 16-dimensional
representation, each one containing five sectors of dimensions  1,4,6,4,1
corresponding respectively
 to 0-forms, 1-forms, 2-forms, 3-forms, and
4-forms.
As explained above, the constraints imply that only two block-sectors made
of one 0-form and
one 1-form sectors are independent wave-functions. The one-form
sector, and the corresponding $4\times 4$ Hamiltonian matrix,
determine the dynamics of spin-one particles.
Moreover, in   a Schwarzschild metric with characteristic radius $r_0$,the
motion is planar, so one can separate the variables and write
\be
\label{eq1}
S= -Et+L\varphi+S_r(r)
\ee
The spin-dependent part of Hamilton-Jacobi's equation
is obtained by the substitution
\be
\label{eq2} p_\m \omega_{ab} ^\m \Sigma ^{ab}={r_0 E \over
r^2}\Sigma ^{01}- {2L \over r^2}\left( 1-{r_0 \over r} \right)^{1/2} \Sigma
^{13}
\ee
where
\be
\Sigma ^{01}={1 \over 2} \pmatrix{ 0&1&0&0\cr -1&0&0&0\cr
0&0&0&0\cr 0&0&0&0}, \quad \Sigma ^{13}={1 \over 2}
\pmatrix{ 0&0&0&0\cr 0&0&0&i\cr
0&0&0&0\cr 0&-i&0&0}
\ee
By inserting (\ref{eq1}) and (\ref{eq2}) into Hamilton-Jacobi's  equation
(\ref{hj}), one
obtains a matricial equation for ${\partial S \over \partial r}$.
The diagonalization can be done easily, and one gets three
possibilities $S_\epsilon$ for the classical action, indexed by
$\epsilon =0,\pm 1$
\be
{E^2 \over c^2}\left( 1-{r_0 \over r}\right)^{-1} - \left(
{L^2 \over r^2}+m^2 \right) -\left( 1-{r_0 \over r}\right) \left(
{\partial S_\epsilon \over \partial r}\right) \nn\\+2\epsilon {\hbar \over
r^2} \left( L^2\left( 1-{r_0 \over r}\right)
-\left( {r_0 E \over 2c}\right )^2
\right)^{1/2} =0
\ee
(We have restored the dependence in the speed of light $c$.)
The energy $E$ and the angular momentum $L$ are constants of motion of the
particle. The values  $\epsilon =0,\pm 1$ correspond to the three
possible projections   of the spin  along a given spatial axis in the
rest frame of the particle. The case $\epsilon=0$ corresponds to the
geodesic trajectory followed by the scalar particle.
Far from the Schwarzschild horizon, we can use the standard
techniques of integration of
Hamilton-Jacobi's equation to determine the three possibilities
for the shift of the perihelion
over a quasi-periodic trajectory.
This amounts to replace $L$ in the classical formulas
\cite{Landau} by an effective angular momentum
$L_\epsilon$ defined by
\be
L_\epsilon^2=
L^2+2\epsilon \hbar L
\sqrt{1-\left({r_0 E\over 2Lc}\right)^2}
\ee
(Notice that near the horizon, unitarity breaks down).
In the case of a massive particle, the shift of the perihelion is
thus given by:
\be
\delta \phi_\epsilon={3 \pi \over 2}
\left({mcr_0 \over L_\epsilon}\right)^2
\sim
\delta \phi_0\left(1-\epsilon {\hbar\over L}
\sqrt{1-\left({r_0 E\over 2Lc}\right)^2} \ \right)
\ee
For a non-relativistic $Z^0$ orbiting quasi-tangentially to the sun at a speed
of $10^5 m/s$, which is approximately the circular velocity around the sun
, we find $ |{{\delta \phi_+ -\delta \phi_0}|/
\delta \phi_0}\sim\hbar/ L\sim 10^{-21}$
, which is much to small to be detected.

The solutions of  Hamilton-Jacobi's equation are continuous when $m\to 0$.
However, in this limit the interpretation of its solution $S$
is different.
The particle is a photon following the laws of the geometrical optics,
$S$ is the eikonal of the light ray,
and  ${\delta S \over \delta X^\m}$  is its wave-vector. The solution
$\epsilon=0$ must then be rejected.
In this massless case, one finds for  the deflections  of the two
helicities $\epsilon=\pm 1$ the
following formula
\be
\delta \phi_\epsilon={2r_0 \omega \over cL_\epsilon}
\sim \delta \phi_0\left(1-\epsilon {\hbar\over 2L}
\sqrt{1-\left({r_0 E\over 2Lc}\right)^2} \ \right)
\ee
where $\omega={E \over \hbar}$.
For an optical photon of wavelength $\lambda =7 \times 10^{-7} m$ (red)
skimming past the sun, we find $ |{{\delta \phi_+ -\delta \phi_0}|/
\delta \phi_0} \sim {\hbar \over 2L}={\lambda \over 2R_{sun}}\sim
10^{-15}$.
(Note that this ratio does not depend on $\hbar$: the gravitational field
interacts classically with the two polarizations of the electromagnetic
field.)
However, this doubling of Einstein's rings is to small to be detected.

\end{document}